# Imaging with flat optics: metalenses or diffractive lenses?

*Sourangsu Banerji,[1] Monjurul Meem,[1] Berardi Sensale-Rodriguez[1] and Rajesh Menon[1,2,a]*

[1]Department of Electrical and Computer Engineering, University of Utah, Salt Lake City, UT 84112, USA.

[2]Oblate Optics, Inc. 13060 Brixton Place, San Diego CA 92130, USA.

[a] rmenon@eng.utah.edu

**ABSTRACT**

Recently, there has been an explosion of interest in metalenses for imaging. The interest is primarily based on their sub-wavelength thicknesses. Diffractive lenses have been used as thin lenses since the late 19$^{th}$ century. Here, we show that multi-level diffractive lenses (MDLs), when designed properly can exceed the performance of metalenses. Furthermore, MDLs can be designed and fabricated with larger constituent features, making them accessible to low-cost, large area volume manufacturing, which is generally challenging for metalenses. The support substrate will dominate overall thickness for all flat optics. Therefore the advantage of a slight decrease in thickness (from ~2$\lambda$ to ~$\lambda$/2) afforded by metalenses may not be useful. We further elaborate on the differences between these approaches and clarify that metalenses have unique advantages when manipulating the polarization states of light.

# INTRODUCTION

Lenses are fundamental to imaging systems. Conventional lenses exploit refraction to focus light [1]. As a result, a fundamental trade-off increases the thickness and weight of optics with increasing numerical aperture (or resolution). As illustrated in Fig. 1a with the example of a simple plano-convex lens, larger bending angles require larger thicknesses. Recently, there has been significant interest in reducing the thickness and weight of lenses by exploiting diffraction. In such "flat-lenses," focusing is achieved by spatially arranging "zones" that impart appropriate phase to achieve constructive interference of the transmitted waves at the focus [2, 3]. As illustrated in Fig. 1b, larger bending angles may be achieved with no change in thickness, simply by decreasing the local period of the diffractive structure. In order to ensure constructive interference, each ray must be locally phase shifted to compensate for the variation in its total optical path length to the focus. In traditional diffractive lenses, this is achieved by engineering the path traversed by the ray within the diffractive lens itself, as illustrated in Fig. 1c. In comparison to travelling the same distance in air, the optical path delay for a thickness, t is $\Delta = (n-1)t$, which then corresponds to a phase shift of $\Delta/\lambda * 2\pi$, where n is the refractive index of the material and $\lambda$ is the wavelength of light. In order to achieve a phase shift of $2\pi$, t must be at least $\lambda/(n-1) \sim 2\lambda$ for n=1.5. It is noted that diffractive lenses with numerical aperture (NA) > 1 under water immersion were demonstrated more than a decade ago [4]. In order to increase the focusing efficiency, blazed or multi-level diffractive lenses (MDLs) were also developed to approximate the optimal continuous phase distribution (see Fig. 1d). In fact, it was widely recognized that close to 100% efficiency could be achieved with such blazed diffractive optics [5]. However, at high numerical apertures, there is a rapid drop in efficiency due to the resonance conditions [6, 7]. It was also quite definitively shown that this drop could be avoided by parametric optimization of the geometry of the constituent structures of the diffractive lens using both simulations [7, 8] and experiments [9, 10]. Another Achilles heel for diffractive lenses has been their poor broadband performance, which was overcome for discrete wavelengths via harmonic phase shifts [11] and by using higher-orders of diffraction [12]. We extended this work to continuous broadband spectra using efficient numerical techniques [13-15] and advanced multi-level nanofabrication at visible [16-20] and Terahertz spectra [21]. *Here, we combine this multi-level approach with parametric optimization to show that high efficiency at high numerical aperture is feasible for both narrowband and broadband operation, which we believe has not been clearly demonstrated before.*

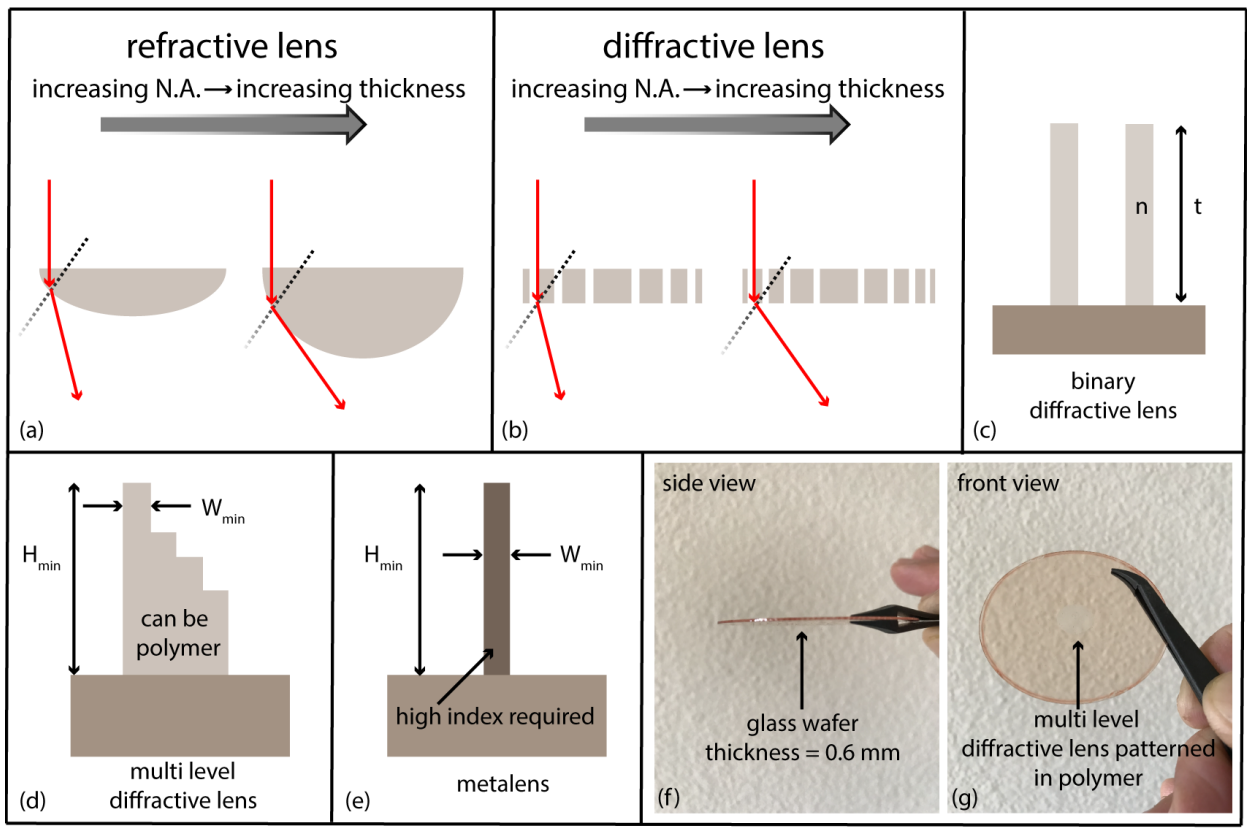

*Figure 1:* Bending of light via (a) refraction and (b) diffraction. Schematic of the constituent element of a (c) conventional binary diffractive lens, (d) multi-level diffractive lens (MDL) and (e) metalens. Photographs of a broadband visible MDL fabricated in a polymer film on a glass

*substrate are shown in (f) side-view emphasizing the small thickness, which is dominated by the substrate and (g) front view.*

Recently, metalenses were proposed as a means to reduce the overall thickness of the conventional diffractive lens to sub-wavelength regimes by exploiting magnified phase changes that can occur in resonators [22-29]. Rather than using traversed path to create a phase shift, appropriately designed subwavelength antenna elements could achieve the same effect (see Fig. 1e). In this report, we show that the advantages of metalenses might be vastly over-stated and that the decrease in thickness from about 2$\lambda$ (achievable via MDLs) to less than $\lambda$ may not be useful for the majority of applications. To emphasize this point, we show a photograph from the side-view (Fig. 1f) of a multi-level diffractive lens that is corrected for the visible spectrum. This device was fabricated in a polymer film atop a glass wafer (thickness~0.6mm) as shown in Fig. 1g [19,20]. We point out that the support substrate will dominate the overall thickness in all cases, and thereby obviate any advantage due to reduction in the device thickness.

We further make the case that MDLs can achieve the same or better optical performance when compared to metalenses. To illustrate this point, we first performed an exhaustive literature survey of metalenses that have been reported so far. A summary of this survey is included in the *Supplementary Information*. Then, we selected exemplary metalenses that operate in the narrowband and in the broadband spectral regimes at low, medium and high numerical apertures, and we designed MDLs having the same optical specifications (focal length, numerical aperture and operating wavelengths). Finally, we compared the focusing efficiencies of the MDLs to those of the corresponding metalenses. Table 1 summarizes the key results. The first 3 columns are the optical specifications. Comparing the focusing efficiencies in columns 6 and 9 confirm that MDLs indeed perform better than metalenses. For the MDLs, we used a commonly available polymer photoresist (Shipley S1813) as the constituent material, since it exhibits high transmission in most wavelength regimes of interest here (measured dispersion is included in the *Supplementary Information*), and we have previously fabricated several MDLs in this material [16-20]. In all cases, we assume unpolarized input light for the MDLs.

*Table 1*: *Summary of performance of MDL and Metalens for same optical specifications. Note that $W_{min}$ and $H_{max}$ are defined in Figs. 1d and 1e for MDL and metalens, respectively.*

| Narrowband | | | Multi-level diffractive lens | | | Metalens | | |
|---|---|---|---|---|---|---|---|---|
| NA | focal length | λ | $W_{min}$ | $H_{max}$ | Efficiency | $W_{min}$ | $H_{max}$ | Efficiency |
| 0.2 | 67μm | 530nm | 1μm | 1.1μm | 93% | 0.05μm | 0.6μm | 92% [25] |
| 0.6 | 200μm | 532nm | 0.4μm | 1.1μm | 90% | 0.25μm | 0.6μm | 87% [26] |
| 0.97 | 25μm | 1550nm | 0.75μm | 3.1μm | 87% | 0.2μm | 0.95μm | 72% [23] |
| Broadband | | | Multi-level diffractive lens | | | Metalens | | |
| NA | focal length | λ | $W_{min}$ | $H_{max}$ | Efficiency | $W_{min}$ | $H_{max}$ | Efficiency |
| 0.2 | 63μm | 470nm to 670nm | 1μm | 2μm | 81% | 0.08μm | 0.6μm | 50% [27] |
| 0.35 | 155μm | 3μm to 5μm | 4μm | 10μm | 86% | 0.4μm | 2μm | 70% [28] |
| 0.81 | 2μm | 560nm to 800nm | 0.35μm | 1.6μm | 70% | 0.055μm | 0.488μm | 69% [29] |

Thirdly, we point out that the fabrication complexity of metalenses is far higher than those of MDLs. As can be seen in Table 1 (columns 4 and 7), the minimum feature widths required for metalenses are significantly smaller than those for MDLs. In addition, metalenses generally require high-index materials (see Tables S1 and S2 in the *Supplementary Information*), whereas MDLs can be fabricated in low-index polymers. It is important to appreciate that any transparent material can be used for the MDL. This allows MDLs to be mass manufactured at low cost via high-volume imprinting techniques [30].

**RESULTS AND DISCUSSION**

Our design methodology involves nonlinear optimization to select the heights of the constituent elements of the MDL in order to maximize focusing efficiency averaged over all wavelengths of interest as described previously [18-20]. In congruence with work in metalenses, we define focusing efficiency as the ratio of the power within a spot of diameter equal to 3 times the simulated full-width at half-maximum (FWHM) to the total incident power [23]. The point-spread function of each MDL was simulated using the finite-difference time-domain method with the incident electric field polarized in the plane of the MDL. Averaging the fields over the two orthogonal polarization directions of the electric field simulates the PSF from unpolarized light. All analysis in the main text utilized this PSF assuming unpolarized input. Details of our simulation are described in the *Supplementary Information*. We note that not all papers follow a consistent method for calculating focusing efficiency. Therefore, we have included a brief description of the methods in the selected metalens papers in the *Supplementary Information*.

*(1) Narrowband MDLs*

First, we consider the design of MDLs for discrete wavelengths (narrowband). Following the parameters from Table 1, we designed 3 MDLs with (focal length, numerical aperture) = (67μm, 0.2), (200μm, 0.6) and (25μm, 0.97).

The optimized designs represented by the height distribution of the concentric rings are illustrated in Figs. 2a, 2c and 2e for NA=0.2, 0.6 and 0.97 MDLs, respectively. The corresponding simulated point-spread functions (PSFs) are shown in Figs. 2b, 2d and 2f, respectively. The FWHM noted in the insets of the PSFs confirm close to diffraction-limited performance. The simulations confirm that even at NA as high as 0.97 efficiencies over 87% are maintained, which are superior to those of corresponding metalenses (Table 1). We note that shadowing effects can clearly impact focusing efficiencies at high NA for both metalenses and MDLs. Our simulations simply point out that metalenses do not offer any advantage over MDLs for narrowband operation, while exhibiting equivalent optical performance.

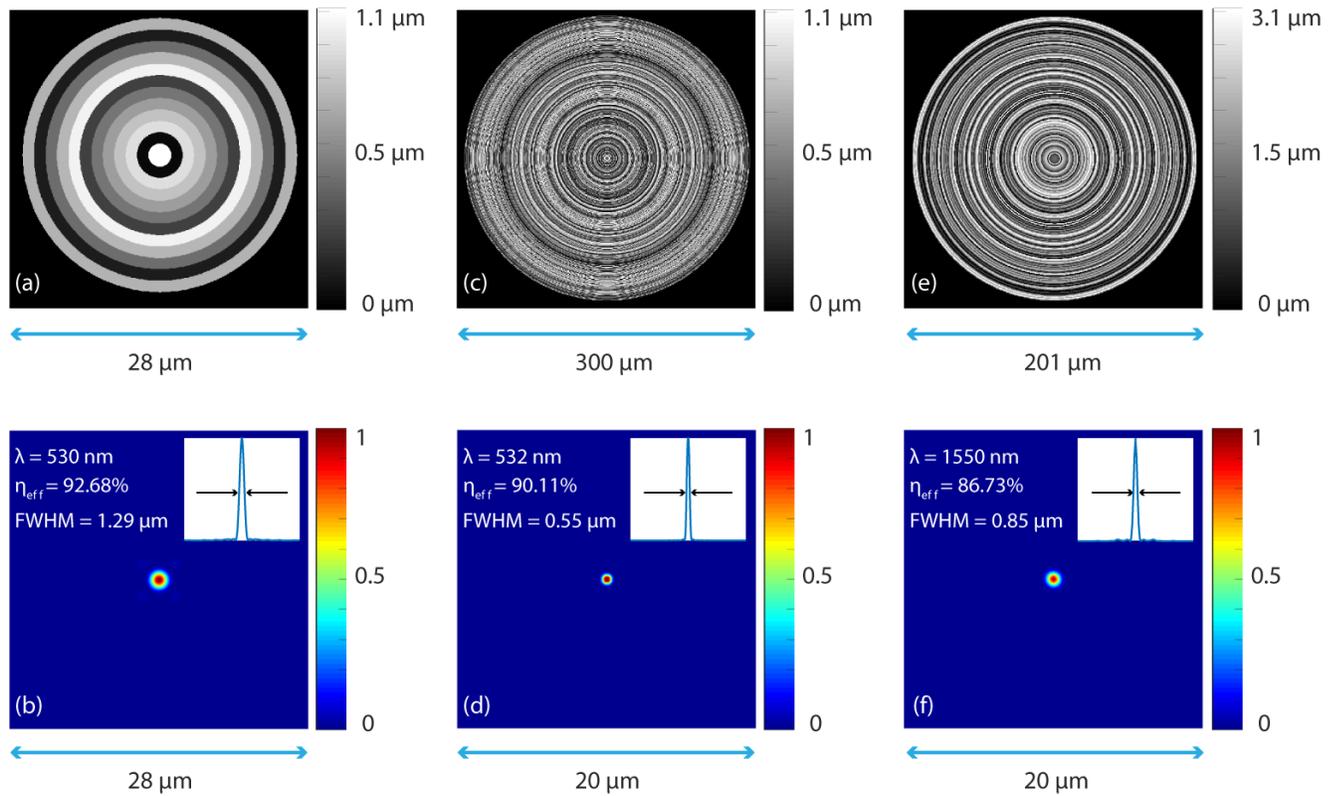

*Figure 2:* Narrowband MDLs. Design (top row) and simulated point-spread function (bottom row) for low (a, b), medium (c, d) and high-NA (e, f) MDLs are shown.

*(2) Broadband MDLs*

One of the big advantages of MDLs as we have pointed out before is their good achromatic performance over broad spectral bands [18-21]. Here, we reiterate this claim by directly comparing MDLs with metalenses of the same optical specifications. Again, following the parameters from Table 1, we designed 3 broadband MDLs with (focal length, numerical aperture) = (63μm, 0.2), (200μm, 0.36) and (2μm, 0.81). The optimized designs represented by the height distribution of the concentric rings are illustrated in Figs. 3a, 3f and 3k for NA=0.2, 0.36 and 0.81 MDLs, respectively. The corresponding simulated point-spread functions (PSFs) for 3 representative wavelengths are shown in Figs. 3b-d, 3g-i and 3i-n, respectively. Again, the FWHM noted in the insets of the PSFs confirm close to diffraction-limited performance for all wavelengths. The simulations confirm that even at NA as high as 0.81 efficiencies of 70% are maintained across the entire band, which are superior to those of corresponding metalenses (Table 1).

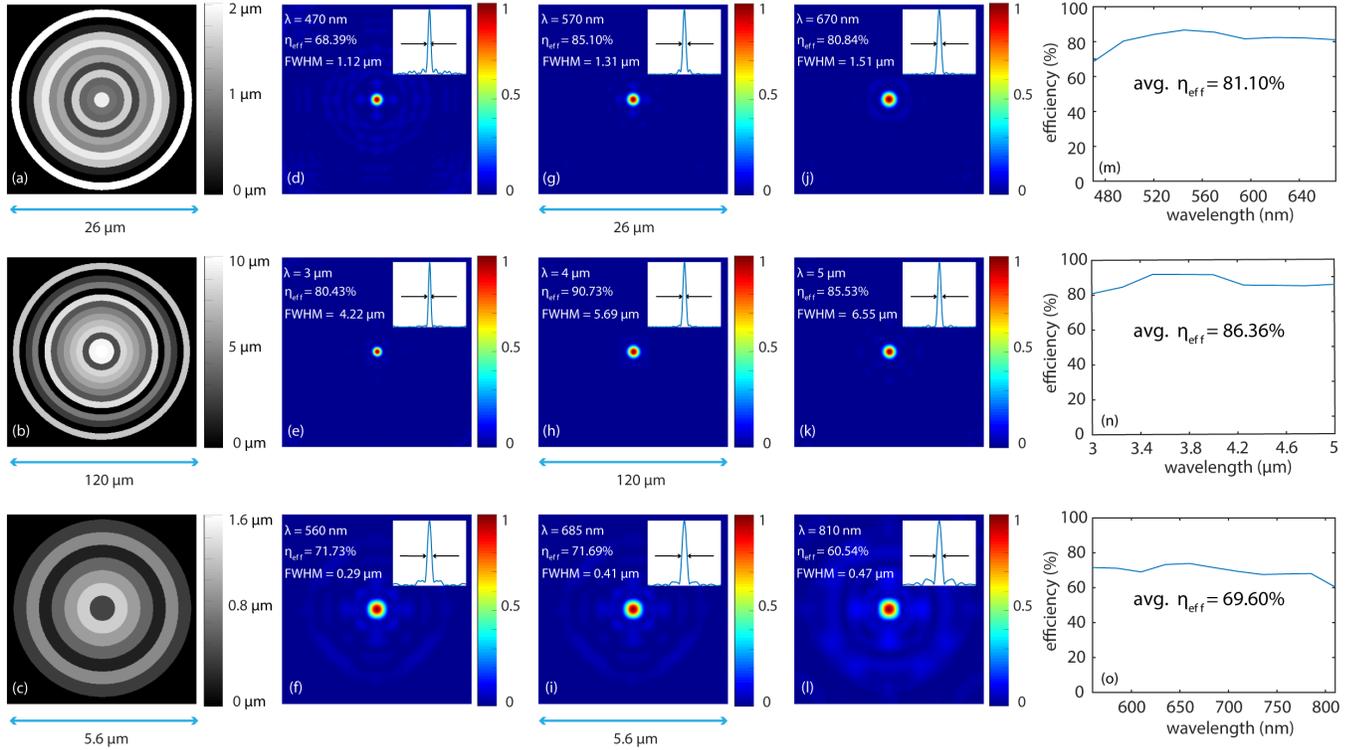

*Figure 3:* *Broadband MDLs. Design (a-c) and simulated point-spread functions (d-l) and simulated focusing efficiency spectra (m-o) for low, medium and high-NA MDLs.*

*(3) Aberrations analysis*

When illuminated by a normally incident uniform plane wave, an ideal lens will generate a perfectly spherical wavefront that converges to the ideal focus. Aberrations in an actual lens are defined as the difference between the actual wavefront from this ideal wavefront. Here, we use the simulated wavefront to analyze the aberrations that are present in MDLs. Using the Zernike-polynomials representation of aberrations, we can calculate the wavefront errors as illustrated in Fig. 4 for the broadband MDL with NA=0.81, f=2μm computed at λ=560nm. Similar results for the other lenses as well as details of the aberrations analysis are included in the *Supplementary Information*.

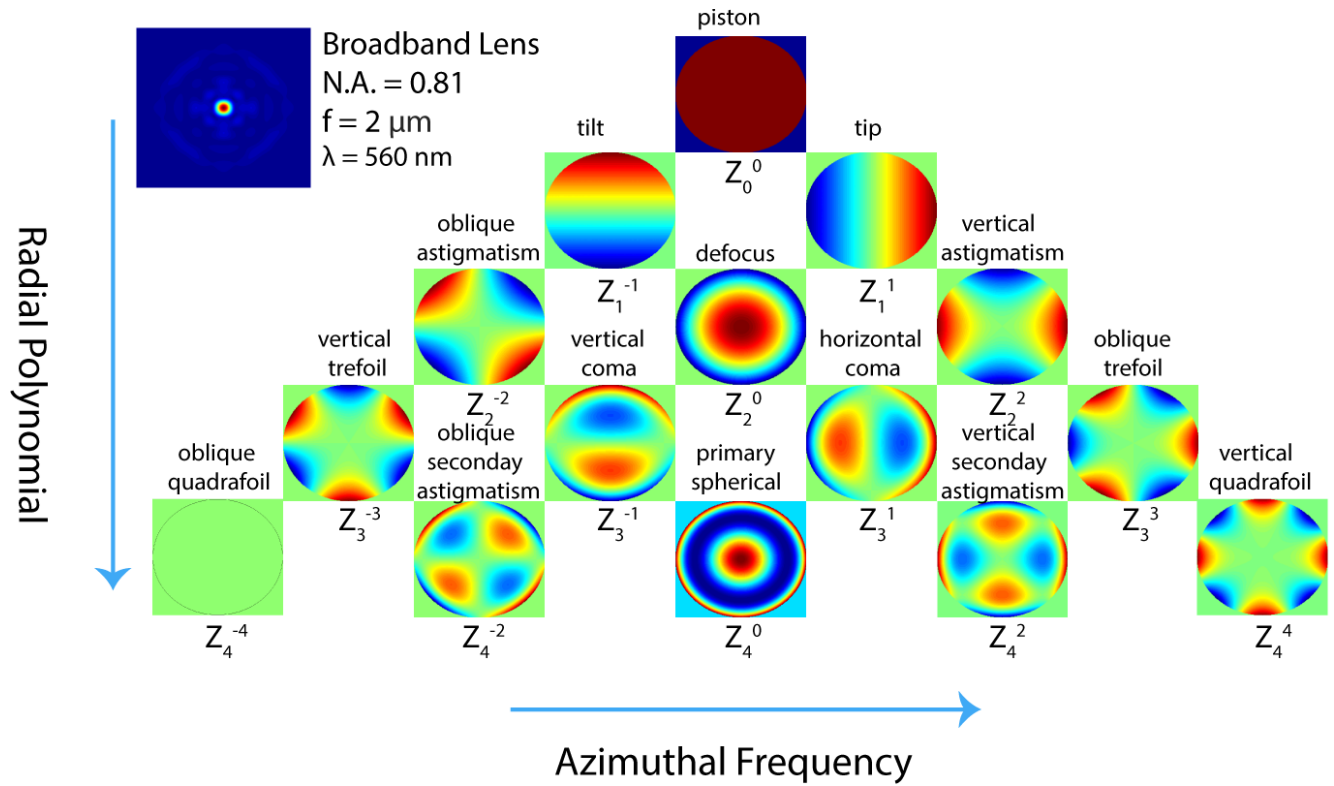

**Figure 4:** *Aberrations-analysis in form of Zernike polynomials for NA=0.81, f=2μm MDL simulated at λ=560nm.*

Furthermore, table 2 summarizes the Zernike coefficients (in units of wavelengths) for the narrowband MDL with NA=0.97, f=25μm, λ=1550nm and the broadband MDL with NA=0.81, f=2μm and simulated at representative wavelengths of λ=560nm, 685nm and 810nm.

These calculations confirm that MDLs have extremely low aberrations and the broadband MDLs exhibit very low variation in aberrations across the operating wavelength range.

Table 2: Zernike coefficients (in units of λ) for two exemplary high-NA MDLs; one narrowband and another broadband.

| Narrowband (N.A. = 0.97) | Multi-level diffractive lens | | | | | | | | | |
|---|---|---|---|---|---|---|---|---|---|---|
| Wavelength | Piston | Tip | Tilt | Defocus | Vertical astigmatism | Horizontal astigmatism | Vertical coma | Horizontal coma | Oblique trefoil | Vertical trefoil |
| 1550 nm | 3.53E-05 | -6.89E-22 | -1.77E-22 | -8.78E-05 | 5.78E-21 | 1.56E-22 | 7.21E-22 | 6.07E-22 | 1.36E-4 | 4.66E-22 |
| Broadband (N.A. = 0.81) | Multi-level diffractive lens | | | | | | | | | |
| Wavelength | Piston | Tip | Tilt | Defocus | Vertical astigmatism | Horizontal astigmatism | Vertical coma | Horizontal coma | Oblique trefoil | Vertical trefoil |
| 560 nm | 1.85E-02 | 1.84E-19 | -6.75E-20 | -2.47E-2 | 3.10E-19 | 8.26E-20 | 2.04E-19 | -9.55E-20 | 4.06E-2 | 1.65E-18 |
| 685 nm | 2.16E-02 | -5.61E-19 | -2.08E-20 | -3.26E-2 | 2.97E-18 | 7.66E-20 | 7.65E-19 | -1.18E-19 | 3.18E-2 | 1.38E-18 |
| 810 nm | 4.23E-02 | -4.21E-19 | 4.96E-20 | -4.56E-2 | -2.68E-18 | 6.08E-20 | 1.90E-18 | 2.11E-19 | 2.21E-2 | 3.57E-18 |

*(4) Where are meta-optics useful?*

Finally, we would like to clarify the regimes where meta-optics (where we include metamaterials, metasurfaces and metalenses) have distinct advantages over MDLs and conventional diffractive optics. Meta-optics have the advantage of extreme form birefringence, which enable them to manipulate the polarization states of light in unique manners. A few illustrative examples of these advantages are in polarimetric imaging [31], high-efficiency polarizers [32] and polarization sensitive optics [33]. Additionally, their sub-wavelength dimensions are extremely useful in integrated optics and photonics, where integration density is a critical parameter for technology adoption [34-36].

**CONCLUSION**

Using a series of rigorous simulations, we conclude that multi-level diffractive lenses, when designed appropriately, can provide better optical performance, while being significantly simpler to manufacture, when compared to metalenses. MDLs can exploit the relatively mature mass manufacturing capabilities that exist in the holograms industry to create low-cost large-area flat optics, enabling a new era of ultra-lightweight, thin optical systems.

**Methods**

All MDL designs were obtained using nonlinear optimization using a modified gradient-descent-based search algorithm that maximized wavelength-averaged focusing efficiency. The PSF simulations were performed using commercially available FDTD software from Lumerical.

**Acknowledgements**

We thank Tom Tiwald from Woollam for assistance with dispersion measurements, and Apratim Majumder for assistance with FDTD modeling. RM and MM acknowledge funding from Office of Naval Research (#N66001-10-1-4065). BSR and SB acknowledge support from NSF CAREER award: ECCS #1351389.

**Author Contributions**

SB performed the design and simulations. MM performed fabrication and analyzed the results. BSR provided design guidance and analyzed the results. RM provided design guidance and analyzed the results. All authors prepared and edited the manuscript.

**Competing Interests Statement**

**Materials and Correspondence**

Correspondence and materials requests should be addressed to RM at [rmenon@eng.utah.edu](mailto:rmenon@eng.utah.edu).

# Imaging with flat optics: metalenses or diffractive lenses?

# Supplementary Information

*Sourangsu Banerji,[1] Monjurul Meem,[1] Berardi Sensale-Rodriguez[1] and Rajesh Menon[1,2,a)]*

[1]Department of Electrical and Computer Engineering, University of Utah, Salt Lake City, UT 84112, USA.

[2]Oblate Optics, Inc. 13060 Brixton Place, San Diego CA 92130, USA.

[a)] rmenon@eng.utah.edu

## 1. Literature Survey

A brief historical review of metalenses is necessary to appreciate its importance with relevance to its counterparts. The concept of such nanostructured sub-wavelength structures is not new. It actually dates back to year 1995-1996 [**1-6**], when the initial demonstrations of graded effective-index artificial-dielectrics for visible frequencies yielded very disappointing results with low measured diffraction efficiencies. The reasons for failure can be attributed to the following reasons. Firstly, the design assumed "adiabatic" effective index gradient [**2, 4**]. Secondly, improper understanding of the relation between local subwavelength gratings and artificial dielectrics [**1-3**]. Lastly, the modeling was challenging as it led to aspect ratios quite difficult to manufacture with the materials and patterning technologies, which operated during that time [**5**].

Fast-forward 15 years later, the field was again revived when a paper was published in Science that revisited Snell's law at the interface between two uniform media with the help of an ultrathin grating composed of metallic nano-antennas etched on the interface [**7**]. Shortly after this, another paper [**8**] was published which controlled the phase using the plasmonic dispersion inside a waveguiding slit in a metal which ultimately led to the focusing of the incident light beam. This was really an important result from the perspective that it showed that due to this resonance occurring at the interface, the phase delay is amplified in contrast to the propagation delay. Therefore, the constraint of having large aspect ratios can be now be considerably relaxed. This possibility was already quite intuitive in the previous paper also; but was very subtle in the presentation to be noticed. The novelty, which underlined in both these two papers, was the fact that graded phases can now be implemented by carefully designed nano-structures specifically; nano-antennas.

To summarize, metalenses unprecedented success can be attributed to three main reasons. The first reason is that these metalenses can control the phase propagation delay through an effective-index modulation leading to a waveguiding effect of the transmitted wavefront. Secondly, these nanostructures can quite effectively also monitor the phase with graded sizes or orientations. Both the reasons combined provide the advantage to have fine spatial sampling with sub-wavelength "unit-cells"; thereby providing a rapid and robust spatial variation of the wrapped phase at the outer zones of the lens. Lastly, the introduction of a resonance delay (occurrence of a plasmonic resonance at the interface leading to amplified phase delay in contrast to the propagation delay) [**9-12**] to implement resonant high-contrast metalenses by combining two resonances, each covering a standard phase range of $\pi$ [**13-15**]. This also relaxed the constraint on having stringent aspect ratios in the designed metalenses. Later on, it was also shown that by using centrosymmetric or rotationally asymmetric structures, full wavefront control could be achieved with a Berry-phase vortex [16]. As an immediate result, many research groups across the world started demonstrating nanostructured metasurfaces having the ability to control the amplitude, phase, polarization, orbital momentum, absorption, reflectance, emissivity of light with high spatial resolution.

Following are some of the game changing publications in the area among many to have appeared in various reputed journals throughout the past decade and has been provided herein to give a perspective of how the field has evolved until today.

**Table S1: Narrowband Metalenses (highlighted designs were used for comparison in Table 1 in main text).**

| Material | Wavelength | N.A. | Focal Length/ Diameter | Simulated Efficiency | Feature Width/Height | Polarization | Reference |
|---|---|---|---|---|---|---|---|
| c-Si | 532 nm | 0.98 | 5.1 um / 50 um | 71% | 20 nm / 500 nm | Circular | [17] |
| TiO2 | 405 nm | 0.8 | 90 um / 240 um | 86% | 40 nm / 600 nm | Circular | [18] |
| TiO2 | 532 nm | 0.8 | 90 um / 240 um | 86% | 40 nm / 600 nm | Circular | [18] |
| TiO2 | 532 nm | 0.8 | 90 um / 240 um | 86% | 40 nm / 600 nm | Circular | [18] |
| TiO2 | 580 nm | 0.94 | 200 um / 1000 um | 79% | > 100 nm / 550 nm | Polarization Insensitive | [19] |
| a-Si | 715 nm | 0.99 | 42 um / 600 um | 88% | 50 nm / 250 nm | Linear | [20] |
| p-Si | 550 nm | 0.43 | 100 um / 96 um | 38% | 100 nm / 100 nm | Linear | [21] |
| **TiO2** | **530 nm** | **0.2** | **67 um / 26.4 um** | **92%** | **50 nm / 600 nm** | **Polarization Insensitive** | **[22]** |
| **a-Si** | **1550 nm** | **0.97** | **25 um / 100um** | **72%** | **200 nm / 950 nm** | **Circular** | **[10]** |
| Au (gold) | 1550 nm | 0.015 | 3 cm / 0.9 mm | 1% | 50nm / 60 nm | Linear | [23] |
| Au (gold) | 1550 nm | 0.075 | 6 cm / 0.9 mm | 1% | 50nm / 60 nm | Linear | [23] |
| Au (gold) | 676 nm | 0.62 | 2.5 um / 4 um | ~10% | 30 nm / 30 nm | Linear | [24] |
| Au (gold) | 676 nm | 0.57 | 5 um / 7 um | ~10% | 30 nm / 30 nm | Linear | [24] |
| Au (gold) | 676 nm | 0.56 | 7 um / 9.4 | ~10% | 30 nm / 30 nm | Linear | [24] |
| | | | um | | | | |
| a-Si | 850 nm | 0.7 | 10 um / 20 um | 93% | 390 nm / 475 nm | Polarization Insensitive | [9] |
| c-Si | 1060 nm | 0.6 | 40 um / 60 um | 89% | 150 nm / 520 nm | Circular | [25] |
| Au (gold) | 800 nm | - | 5.7 um / 1.06 um | Qualitative Agreement | 34 nm / 50 nm | Circular | [26] |
| Au (gold) | 740 nm | - | 10 um / 8 um | Qualitative Agreement | 50 mm / 40 nm | Circular | [27] |
| TiO2 | 532 nm | 0.6 | 200 um / 300 um | > 87% | 250 nm / 600 nm | Polarization Insensitive | [28] |
| TiO2 | 405 nm | 0.6 | 200 um / 300 um | > 87% | 250 nm / 600 nm | Polarization Insensitive | [28] |
| TiO2 | 660 nm | 0.6 | 200 um / 300 um | > 87% | 250 nm / 600 nm | Polarization Insensitive | [28] |
| TiO2 | 532 nm | 0.85 | 90 um / 300 um | 83% | 250 nm / 600 nm | Polarization Insensitive | [28] |
| TiO2 | 405 nm | 0.85 | 90 um / 300 um | 79% | 250 nm / 600 nm | Polarization Insensitive | [28] |
| TiO2 | 660 nm | 0.85 | 90 um / 300 um | 84% | 250 nm / 600 nm | Polarization Insensitive | [28] |
| a-Si | 1550 nm | 0.2 | 50 mm / 20 mm | > 90% | 830 nm / 600mm | Polarization Insensitive | [29] |
| PbTe | 5.2 um | - | 5.2um / - | - | 2500 nm / 650 nm | - | [30] |
| GaSb | 4 um | 0.36 | 155 um / ~300 um | 80% | 30 um / 2 um | Polarization Insensitive | [31] |
| a-Si | 4 um | 0.45 | 300 um / 300 um | ~ 96% | 600 nm / 2 um | Polarization Insensitive | [32] |
| a-Si | 633 nm | 1 | 633 nm / 30 um | 2.50% | 22 nm / 0.24 um | Linear | [33] |
| Si | 473 nm | 0.6294 | 10 um / - | 21.13% | 42 nm / 400 nm | - | [34] |
| Si | 532 nm | 0.6294 | 10 um / - | 54.66% | 67 nm / 400 nm | - | [34] |
| Si | 632.8 nm | 0.6294 | 10 um / - | 31.49% | 102 nm / 400 nm | - | [34] |
| Ag/Alumina/Ag | 633 nm | - | 2.2 um / 3.6 um | Qualitative Agreement | - / 1332 nm | Polarization Insensitive | [35] |
| Ag/Alumina/Ag | 633 nm | - | 1.9 um / 2.8 um | Qualitative Agreement | - / 851 nm | Polarization Insensitive | [35] |
| GaN | 430 nm | 0.22 | 110 um / 50 um | - | 50 nm / 600 nm | Circular | [36] |
| GaN | 532 nm | 0.22 | 110 um / 50 um | - | 70 nm / 600 nm | Circular | [36] |
| GaN | 633 nm | 0.22 | 110 um / 50 um | - | 100 nm / 600 nm | Circular | [36] |
| Ag / TiO2 | 660 nm | - | 1.4 um / - | - | - / 2.8 um | - | [37] |
| Ag/Alumina | 830 nm | - | 1.8 um / - | - | - / 2 um | - | [37] |

| Material | Wavelength | N.A. | Focal Length/Diameter | Simulated Efficiency | Feature Width/Height | Polarization | Reference |
|---|---|---|---|---|---|---|---|
| TiO2 | 633 nm | 0.37 | 500 um / 400 um | ~ 100% | 300 nm / 155 nm | - | [38] |
| TiO2 | 633 nm | 0.89 | 100 um / 400 um | ~ 100% | 300 nm / 155 nm | - | [38] |
| Ag | 365 nm | - | 1 um / 1.8 um | - | 10 nm to 20 nm / 200 nm | Linear | [39] |
| Al | 365 nm | - | 0.5 um / 1.8 um | - | 10 nm to 20 nm / 200 nm | Linear | [39] |
| Ag | 365 nm | - | 0.5 um / 1.8 um | - | 10 nm to 20 nm / 200 nm | Linear | [39] |
| PbTe | 5.2 um | 0.71 | 500 um / - | ~ 80% | 2500 nm / 650 nm | Linear | [40] |
| Au (gold) | 600 nm | 0.58 to 0.85 | 3 um, 5um, 7 um / 10 um | ~20% | 100 nm, 60 nm, 40 nm / 40 nm | Linear | [41] |
| Au (gold) | 785 nm | 0.58 to 0.85 | 3 um, 5um, 7 um / 10 um | ~20% | 100 nm, 60 nm, 40 nm / 40 nm | Linear | [41] |
| Au (gold) | 980 nm | 0.58 to 0.85 | 3 um, 5um, 7 um / 10 um | ~20% | 100 nm, 60 nm, 40 nm / 40 nm | Linear | [41] |
| a-Si | 658 nm | 0.3511 | 400 um / 300 um | Qualitative Agreement | - / 15 to 50 nm | Polarization Insensitive | [42] |
| Si3N4 | 633 nm | 0.98 | 10 um / 100 um | Qualitative Agreement | - / 695 nm | Unpolarized | [43] |
| Si3N4 | 633 nm | 0.78 | 4 mm / 1 cmm | Qualitative Agreement | - / 695 nm | Unpolarized | [43] |
| Ag (Silver) | 1550 nm | - | 10 um / 8 um | - | 80 nm / 50 nm | Linear | [44] |
| TiO2 | 633 nm | upto 0.8 | 2 um to 14 um / 5.4 um | 88.50% | 10 nm - 150 nm / 488 nm | Polarization Insensitive | [45] |
| TiO2 | 450 nm | 0.1 | 1 mm / - | - | 320 nm / 90 nm | Polarization Insensitive | [46] |
| TiO2 | 532 nm | 0.1 | 1 mm / - | - | 320 nm / 90 nm | Polarization Insensitive | [46] |
| TiO2 | 633 nm | 0.1 | 1 mm / - | - | 320 nm / 90 nm | Polarization Insensitive | [46] |
| TiO2 | 800 nm | 0.247 | 40 um / 20 um | 99% (cylindrical lens) | 90 nm - 200 nm / 250 nm | - | [47] |

**Table S2: Broadband Metalenses (highlighted designs were used for comparison in Table 1 in main text).**

| Material | Wavelength | Bandwidth | N.A. | Focal Length/Diameter | Simulated Efficiency | Feature Width/Height | Polarization | Reference |
|---|---|---|---|---|---|---|---|---|
| a-Si | 1.3 um - 1.65 um | 350 nm | 0.24 | 200 um / 100 um | - | 100 nm / 1400 nm | Polarization Insensitive | [48] |
| a-Si | 1.2 um - 1.65 um | 450 nm | 0.13 | 800 um / 200 um | - | 100 nm / 800 nm | Polarization Insensitive | [48] |
| a-Si | 1.2 um - | 200 nm | 0.88 | 30 um / | - | 100 nm / 800 | Polarization | [48] |
| | 1.40 um | | | 100 um | | nm | Insensitive | |
| GaN | 400 nm - 660 nm | 260 nm | 0.106 | 235 um/ 50 um | average 40% (measured) | 45 nm/800 nm | Circular | [49] |
| TiO2 | 460 nm - 700 nm | 240 nm | 0.2 | 67um /26.4um | | 50 nm /600 nm | Polarization Insensitive | [22] |
| a-Si | 1300 nm - 1800 nm | 500 nm | 0.04 | 7.5 mm / 600 um | 24%, 22%, and 28% | 75 nm/ 600 nm | Linear | [50] |
| Au/SiO2/Au | 1.2 nm - 1.68 um | 480 nm | 0.268 | 100 um / 55.55 um | 12.44 % (measured) | 40nm /30 nm | Circular | [51] |
| Au/SiO2/Au | 1.2 nm - 1.68 um | 481 nm | 0.217 | - | 8.4 % (measured) | 40nm /30 nm | Circular | [51] |
| Au/SiO2/Au | 1.2 nm - 1.68 um | 482 nm | 0.324 | - | 8.56 % (measured) | 40nm /30 nm | Circular | [51] |
| TiO2 | 470 nm - 670 nm | 200 | 0.2 | 63 um / 25.2 um | 50% | 80 nm / 600 nm | Circular | [52] |
| PbTe | 5.11 um - 5.29 um | 180 nm | 0.5 mm / - | - | - | / 650 nm | - | [30] |
| GaSb | 3 um - 5 um | 2 um | 0.35 | 155 um / >300 um | 70% | 30 um / 2 um | Polarization Insensitive | [31] |
| a-Si | 5 um - 8 um | 3 um | 0.35 | 30*lambda / - | - | - / 1.5*lambda | Polarization Insensitive | [53] |
| a-Si | 3.7 um - 4.2 um | 0.5 um | 0.45 | 300 um / 300 um | ~ 96% | / 2 um | Polarization Insensitive | [32] |
| GaN | 435 nm - 685 nm | 250 nm | 0.17 | 20 um / 7 um | 50% - 78% | 160nm or 240 nm / 400 nm | Linear | [54] |
| Photoresist ( polymer-ZEP520A) | 436 nm - 685 nm | 250 nm | 0.17 | 20 um / 7 um | 50% - 78% | 160nm or 240 nm / 400 nm | Linear | [54] |
| PbTe | 5.11 um - 5.29 um | 180 nm | - | 0.5 mm / 1 mm | ~ 75% | 2500 nm / 650 nm | Linear | [40] |
| Au | 532 nm - 1080 nm | 548 nm | - | 7 um / 10 um | ~ 20% | 100 nm , 60 nm, 40 nm / 40 nm | Linear | [55] |
| a-Si | 470 nm - 658 nm | 188 nm | 0.3511 | 400 um / 300 um | Qualitative Agreement | - / 15 to 50 nm | Polarization Insensitive | [42] |
| TiO2 | 490 nm - 550 nm | 60 nm | 0.2 | 485 um / 200 um | Qualitative Agreement | varied / 180 nm | - | [56] |
| TiO2 | 560 nm - 800 nm | 240 nm | upto 0.8 | 2 um to 14 um / 5.4 um | > 50% | 10 nm - 150 nm / 488 nm | Polarization Insensitive | [45] |
| fused Si | 486 nm - 656 nm | 170 nm | 0.1 | 100 mm / 20 mm | - | 1300 nm / 560 nm | Polarization Insensitive | [57] |

## 2. Measured Dispersion of Photopolymer S1813

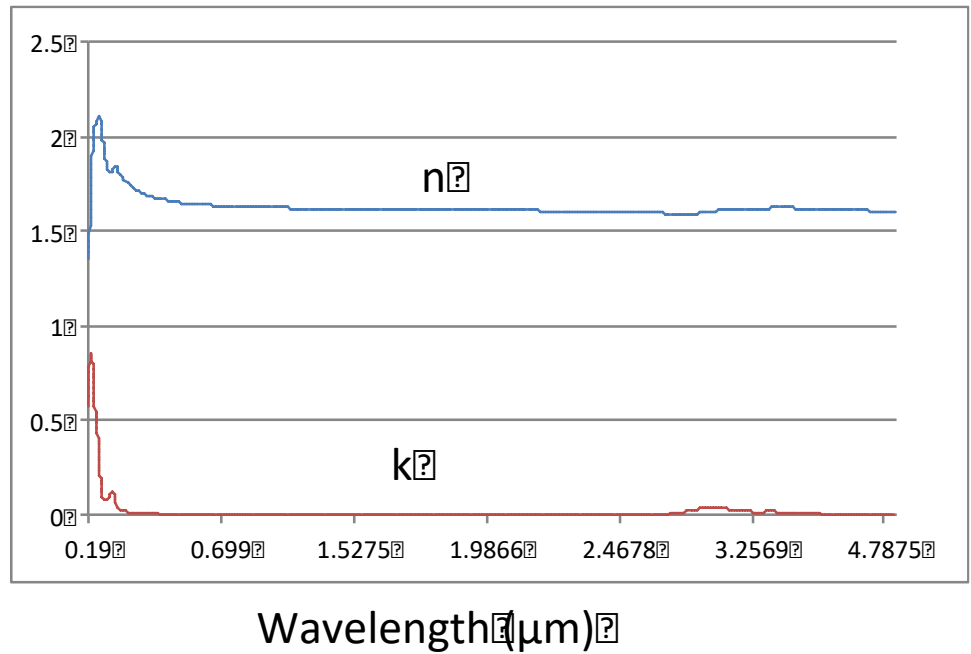

**Figure S1:** Measured dispersion of S1813.

## 3. FDTD Simulations

The full wave FDTD simulations were carried out using Lumerical FDTD Solutions. The material properties (refractive index and absorption coefficient as a function of frequency) of the Photoresist (S1813) was imported into Lumerical directly as the structure's optical data. A ".lsf" script was written to replicate the lens geometry using the same dimensions, which was specified during the optimization process as depicted in **Fig. S2 (a)**. An incident plane wave (type: diffracting [see link]) along the backward "y-axis" direction (with TM polarization) were used to illuminate the diffractive lens surface.

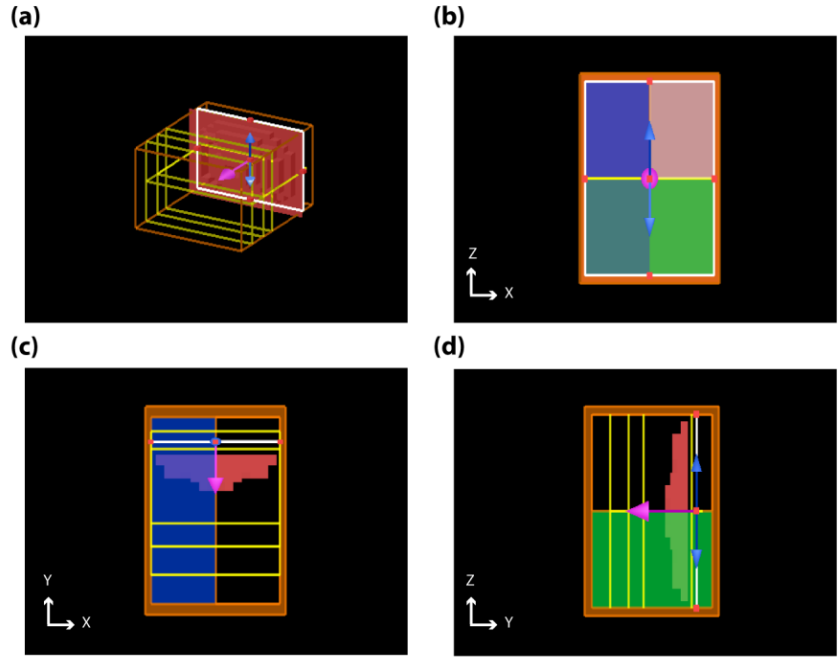

**Figure S2 (a)** Perspective view of the FDTD simulation setup of the spherical lens. **(b)** Visualization of boundary conditions imposed during the simulation. A boundary condition of **(c)** "symmetric" set to x-min and **(d)** "anti-symmetric" set to z-min.

For the broadband excitation, the entire range or bandwidth of the pulse was defined for the appropriate design. The entire FDTD simulation region was considered from the back surface of the spherical lens right up to 1.5 times the distance from the focal plane. A Perfectly Matched Layer (PML) boundary condition set up in the x-max, y-max and both z-min and z-max directions. As seen from **Fig. S2 (b-d)** that due to the inherent symmetry of the designed structure, the x-min boundary was set to "symmetric" and the z-min boundary was set to "anti-symmetric" which reduced the requirements by ¼ of the original simulation requirements in terms of both time and memory. We would like to emphasis here that we tried to impose radial symmetry but could not as it has not yet been made available in the software [see link].

The default mesh was used to simulate the structures instead of a very fine mesh to avoid the huge computation time. The mesh accuracy was kept at "3" which has a good tradeoff for precision and accuracy versus the time and memory requirement. Field monitors placed at different planes above the lens and along the vertical plane to observe the field profiles of the propagating electromagnetic radiation.

**4. Methods of calculating focusing efficiency in selected metalens references**

Reference 25 from main text: W. T. Chen, A.Y. Zhu, J. Sisler, Z. Bharwani, and F. Capasso, "A Broadband achromatic polarization-insensitive metalens consisting of anisotropic nanostructures," arXiv preprint arXiv:1810.05050 (2018).

This paper does not explicitly define the focusing efficiency of the metalens.

Reference 26 from main text: F. Aieta, P. Genevet, P., M.A. Kats, N. Yu, R. Blanchard, Z. Gaburro, and F. Capasso, "Polarization insensitive metalenses at visible wavelengths," Nano Letts, 12(9), pp.4932-4936 (2012).

This paper defines focusing efficiency as "the ratio of the optical power in the focal spot area (circle of radius $2 \times$ FWM spanning the center of the focal spot) to the incident optical power." We computed the focusing efficiency of the corresponding MDL with this definition and confirmed that the number (89.3%) was almost identical to the number with our definition in the main text (90%).

Reference 23 from main text: A. Arbabi, Y. Horie, A. J. Ball, M. Bagheri, and A. Faraon, "Subwavelength-thick lenses with high numerical apertures and large efficiency based on high-contrast transmitarrays," Nature Commun. 6, 7069 (2015).

This paper defines focusing efficiency as the ratio of the optical power in an area of diameter 3 X FWHM to the total incident power (which is the same as what we used).

Reference 27 from main text: W. T. Chen, A. Y. Zhu, V. Sanjeev, M. Khorasaninejad, Z. Shi, E. Lee, and F. Capasso, "A broadband achromatic metalens for focusing and imaging in the visible," Nature Nanotechnology, 13, p.220 (2018).

This paper does not clearly define the definition of focusing efficiency. From the supplementary information, we can surmise that they are using the ratio of the power within the airy disk (which is approximately 2.5 times FWHM) to the total incident power.

Reference 28 from main text: S. Zhang, A. Soibel, S. A. Keo, D. Wilson, B. Rafol, D. Z. Ting, A. She, S. D. Gunapala and F. Capasso, "Solid-immersion metalenses for infrared focal-plane arrays," Appl. Phys. Lett. 113, 111104 (2018).

This paper defines "The focusing efficiency was defined as the optical power over the pixel size of 10 μm at the focus divided by the incident power over the pixel pitch size of 30 μm." This 10μm corresponds to approximately 2 times FWHM of the middle wavelength (4μm).

Reference 29 from main text: Y. Liang, H. Liu, F. Wang, F., H. Meng, J. Guo, J., J. Li, and Z. Wei, "High-efficiency, near-diffraction limited, dielectric metasurface lenses based on crystalline titanium dioxide at visible wavelengths," Nanomaterials, 8, pp: 288 (2018).

From Fig. 6d of this paper, we estimate that the efficiency is defined as power within width of about 3 times FWHM divided by total incident power. The actual definition is not clearly stated.

## 5. Simulated Point Spread Functions (PSFs) [Broadband Lenses]

The simulated point spread functions (PSFs) for the broadband lenses designed at low, mid and high N.A. are provided.

**Low NA Broadband Lens: [NA = 0.2 (470 nm – 670 nm)]**

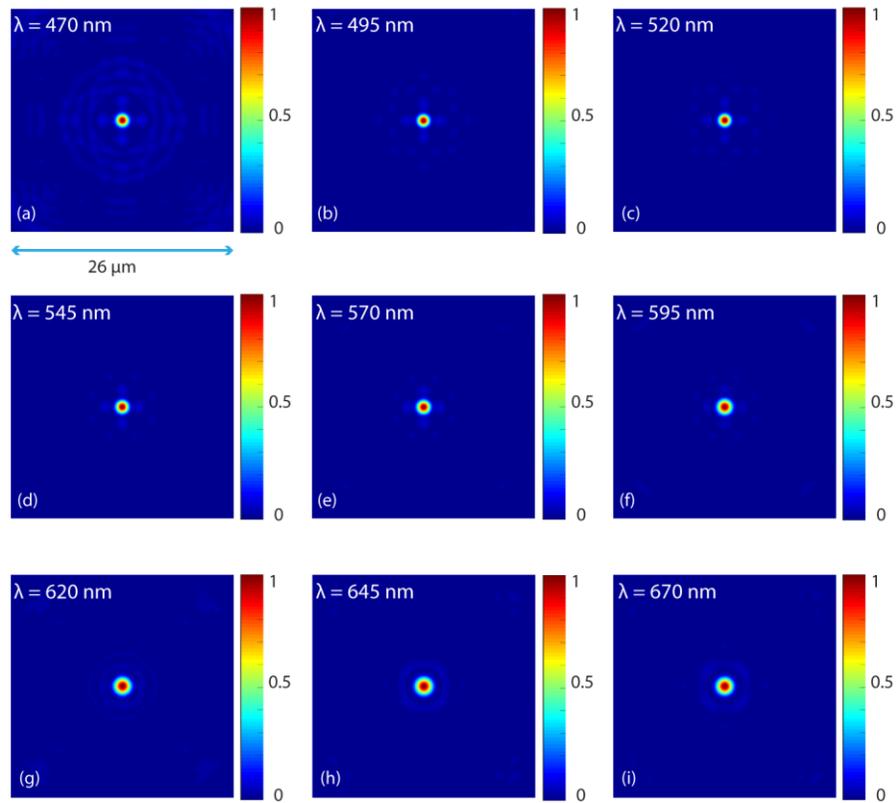

**Figure S3** PSF plot corresponding to a wavelength of **(a)** 470 nm **(b)** 495 nm **(c)** 520 nm **(d)** 545 nm **(e)** 570 nm **(f)** 595 nm **(g)** 620 nm **(h)** 645 nm and **(i)** 670 nm.

**Mid NA Broadband Lens: [NA = 0.36 (3 um – 5 um)]**

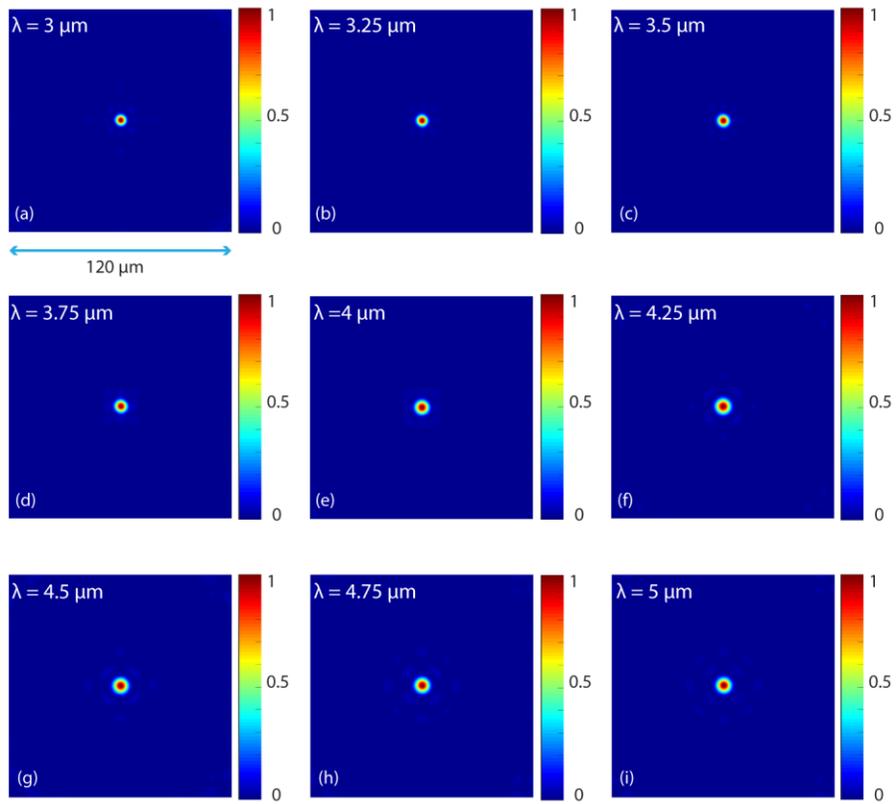

**Figure S4** PSF plot corresponding to a wavelength of **(a)** 3 um **(b)** 3.25 um **(c)** 3.5 um **(d)** 3.75 um **(e)** 4 um **(f)** 4.25 um **(g)** 4.5 um **(h)** 4.75 um and **(i)** 5 um.

**High NA Broadband Lens: [NA = 0.81 (560 nm – 810 nm)]**

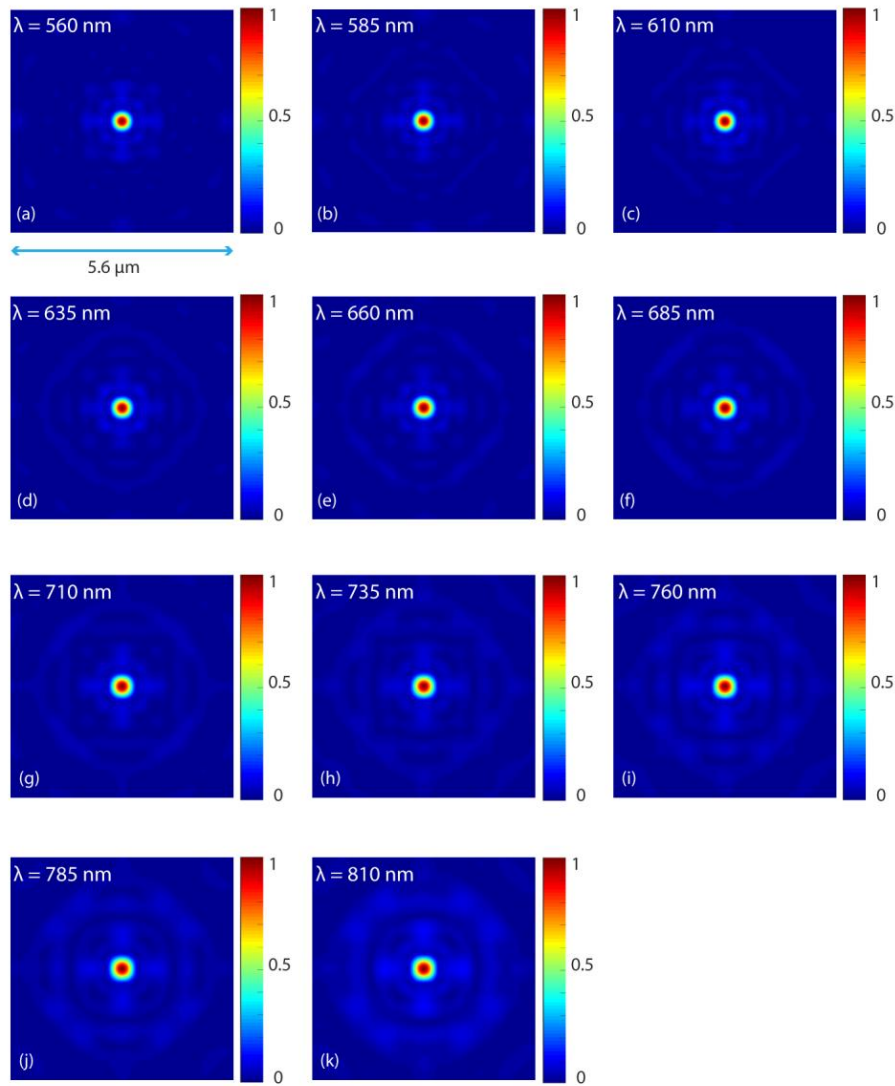

**Figure S5** PSF plot corresponding to a wavelength of **(a)** 560 nm **(b)** 585 nm **(c)** 610 nm **(d)** 635 nm **(e)** 660 nm **(f)** 685 nm **(g)** 710 nm **(h)** 735 nm **(i)** 760 nm **(j)** 785 nm and **(k)** 810 nm.

**6. Simulated FWHM and efficiency**

**Table S3: Low NA Narrowband Lens: [NA = 0.2 (530 nm)]**

| Wavelength (nm) | F.W.H.M. (Diffraction Limit) (um) | F.W.H.M. (Simulation-FDTD) (um) | Efficiency (%) |
|---|---|---|---|
| 530 | 1.2956 | 1.2932 | 92.68 |

**Table S4: Mid NA Narrowband Lens: [NA = 0.6 (532 nm)]**

| Wavelength (nm) | F.W.H.M. (Diffraction Limit) (um) | F.W.H.M. (Simulation-FDTD) (um) | Efficiency (%) |
|---|---|---|---|
| 532 | 0.4433 | 0.5458 | 90.11 |

**Table S5: High NA Narrowband Lens: [NA = 0.97 (1550 nm)]**

| Wavelength (nm) | F.W.H.M. (Diffraction Limit) (um) | F.W.H.M. (Simulation-FDTD) (um) | Efficiency (%) |
|---|---|---|---|
| 1550 | 0.7986 | 0.8458 | 86.73 |

**Table S6: Low NA Broadband Lens: [NA = 0.2 (470 nm – 670 nm)]**

| Wavelength (nm) | F.W.H.M. (Diffraction Limit) (um) | F.W.H.M. (Simulation-FDTD) (um) | Efficiency (%) |
|---|---|---|---|
| 470 | 1.1628 | 1.1244 | 68.39 |
| 495 | 1.2247 | 1.1376 | 80.1 |
| 520 | 1.2865 | 1.1526 | 83.82 |
| 545 | 1.3484 | 1.2318 | 86.31 |
| 570 | 1.4103 | 1.3054 | 85.1 |
| 595 | 1.4721 | 1.454 | 81.35 |
| 620 | 1.534 | 1.5262 | 82.14 |
| 645 | 1.5958 | 1.5738 | 81.84 |
| 670 | 1.6577 | 1.5124 | 80.84 |

**Table S7: Mid NA Broadband Lens: [NA = 0.36 (3 um – 5 um)]**

| Wavelength (um) | F.W.H.M. (Diffraction Limit) (um) | F.W.H.M. (Simulation-FDTD) (um) | Efficiency (%) |
|---|---|---|---|
| 3 | 4.1552 | 4.218 | 80.43 |
| 3.25 | 4.5015 | 4.509 | 83.95 |
| 3.5 | 4.8477 | 4.821 | 90.93 |
| 3.75 | 5.194 | 5.086 | 90.88 |
| 4 | 5.5403 | 5.688 | 90.73 |
| 4.25 | 5.8865 | 6.124 | 85.09 |
| 4.5 | 6.2328 | 6.384 | 85.01 |
| 4.75 | 6.5791 | 5.977 | 84.73 |
| 5 | 6.9253 | 6.546 | 85.53 |

**Table S8: High NA Broadband Lens: [NA = 0.81 (560 nm – 810 nm)]**

| Wavelength (nm) | F.W.H.M. (Diffraction Limit) (um) | F.W.H.M. (Simulation-FDTD) (um) | Efficiency (%) |
|---|---|---|---|
| 560 | 0.3441 | 0.2912 | 71.73 |
| 585 | 0.3595 | 0.3645 | 71.32 |
| 610 | 0.3748 | 0.3578 | 69.22 |
| 635 | 0.3902 | 0.3945 | 73.46 |
| 660 | 0.4055 | 0.3952 | 74.03 |
| 685 | 0.4209 | 0.4104 | 71.69 |
| 710 | 0.4363 | 0.4134 | 69.5 |
| 735 | 0.4516 | 0.4278 | 67.73 |
| 760 | 0.467 | 0.4378 | 68.06 |
| 785 | 0.4823 | 0.4784 | 68.29 |
| 810 | 0.4977 | 0.4745 | 60.54 |

## 7. Aberrations Analysis

The Zernike polynomial coefficient was fitted over a circular shaped pupil. The calculation was done using the reference [**57**]. The fit was achieved with a least squares fit method. The indexing scheme used was Fringe.

**Table S9: Aberrations coefficients**

| Radial degree (n) | Azimuthal degree (m) | Fringe index (j) | Classical name |
|---|---|---|---|
| 0 | 0 | 1 | piston |
| 1 | 1 | 2 | tip |
| 1 | -1 | 3 | tilt |
| 2 | 0 | 4 | defocus |
| 2 | 2 | 5 | vertical astigmatism |
| 2 | -2 | 6 | oblique astigmatism |
| 3 | 1 | 7 | horizontal coma |
| 3 | -1 | 8 | vertical coma |
| 4 | 0 | 9 | primary spherical |
| 3 | 3 | 10 | oblique trefoil |
| 3 | -3 | 11 | vertical trefoil |
| 4 | 2 | 12 | vertical secondary astigmatism |
| 4 | -2 | 13 | oblique secondary astigmatism |
| 4 | 4 | 14 | vertical quadrafoil |
| 4 | -4 | 15 | oblique quadrafoil |

The following lists all the fitting coefficients for the designed lenses:

### Table S10: Low NA Narrowband Lens: [NA = 0.2 (530 nm)]

| Wavelength (nm) | Piston | Tip | Tilt | Defocus | Vertical astigmatism | Oblique astigmatism | Horizontal coma | Vertical coma | Primary spherical | Oblique trefoil | Vertical trefoil | Vertical secondary astigmatism | Oblique secondary astigmatism | Vertical quadrafoil | Oblique quadrafoil |
|---|---|---|---|---|---|---|---|---|---|---|---|---|---|---|---|
| 530 | 4.27E-03 | -4.90E-20 | 1.03E-20 | -0.009314435145 | -1.39E-19 | -1.72E-21 | -9.63E-21 | 1.19E-20 | 0.01542385988 | 1.79E-19 | -2.64E-21 | 8.38E-20 | -6.65E-21 | 2.25E-04 | 8.12E-21 |

### Table S11: Mid NA Narrowband Lens: [NA = 0.6 (532 nm)]

| Wavelength (nm) | Piston | Tip | Tilt | Defocus | Vertical astigmatism | Oblique astigmatism | Horizontal coma | Vertical coma | Primary spherical | Oblique trefoil | Vertical trefoil | Vertical secondary astigmatism | Oblique secondary astigmatism | Vertical quadrafoil | Oblique quadrafoil |
|---|---|---|---|---|---|---|---|---|---|---|---|---|---|---|---|
| 532 | 1.64E-04 | -5.78E-20 | 6.19E-21 | 5.70E-05 | 1.12E-19 | 5.07E-21 | 4.55E-20 | -6.73E-21 | 7.97E-05 | -1.74E-20 | -3.16E-20 | -7.17E-20 | 5.45E-21 | -9.13E-08 | 8.18E-22 |

### Table S12: High NA Narrowband Lens: [NA = 0.97 (1550 nm)]

| Wavelength (nm) | Piston | Tip | Tilt | Defocus | Vertical astigmatism | Oblique astigmatism | Horizontal coma | Vertical coma | Primary spherical | Oblique trefoil | Vertical trefoil | Vertical secondary astigmatism | Oblique secondary astigmatism | Vertical quadrafoil | Oblique quadrafoil |
|---|---|---|---|---|---|---|---|---|---|---|---|---|---|---|---|
| 1550 | 3.53E-05 | -6.89E-22 | -1.77E-22 | -8.78E-05 | 5.78E-21 | 1.56E-22 | 7.21E-22 | 6.07E-22 | 0.0001360991827 | 4.66E-22 | 2.51E-23 | -1.39E-21 | 2.86E-22 | 1.93E-08 | 5.05E-23 |

### Table S13: Low NA Broadband Lens: [NA = 0.2 (470 nm – 670 nm)]

| Wavelength (nm) | Piston | Tip | Tilt | Defocus | Vertical astigmatism | Oblique astigmatism | Horizontal coma | Vertical coma | Primary spherical | Oblique trefoil | Vertical trefoil | Vertical secondary | Oblique secondary | Vertical quadrafoil | Oblique quadra |
|---|---|---|---|---|---|---|---|---|---|---|---|---|---|---|---|

| Wavelength (nm) | Piston | Tip | Tilt | Defocus | Vertical astigmatism | Oblique astigmatism | Horizontal coma | Vertical coma | Primary spherical | Oblique trefoil | Vertical trefoil | Vertical secondary astigmatism | Oblique secondary astigmatism | Vertical quadrafoil | Oblique quadrafoil |
|---|---|---|---|---|---|---|---|---|---|---|---|---|---|---|---|
| 470 | 1.49E-02 | -6.67E-19 | -8.83E-21 | -0.01648955077 | 2.78E-18 | 2.16E-21 | 5.41E-19 | 5.00E-20 | 0.01790526432 | 6.14E-19 | -6.00E-20 | 8.05E-18 | 3.07E-20 | -0.001604880073 | -2.79E-21 |
| 495 | 7.24E-03 | -1.40E-19 | 9.34E-22 | -0.01189857301 | -3.11E-19 | -2.29E-21 | 3.18E-21 | -1.06E-20 | 0.01627985935 | 3.60E-19 | 1.59E-20 | -8.59E-19 | 1.77E-20 | -0.0007075870624 | 7.38E-21 |
| 520 | 6.11E-03 | -2.23E-19 | 3.93E-21 | -0.0109625027 | 1.02E-19 | -1.69E-20 | -2.21E-19 | 1.25E-20 | 0.01720543958 | 3.88E-19 | -1.53E-20 | -2.41E-19 | -1.09E-20 | -0.0016137652 | -1.40E-20 |
| 545 | 5.76E-03 | -1.15E-19 | 5.19E-21 | -0.01036614896 | -4.39E-19 | 2.67E-20 | -2.70E-19 | 4.25E-20 | 0.01825712339 | 9.85E-20 | 8.80E-21 | -9.94E-20 | -1.33E-19 | -0.002448728824 | 3.77E-20 |
| 570 | 6.34E-03 | 8.19E-20 | -6.06E-20 | -0.01090366043 | 4.00E-19 | -4.77E-20 | 2.23E-19 | 6.43E-20 | 0.01995126743 | 2.57E-19 | -6.13E-21 | -1.95E-19 | 9.59E-20 | -0.002923797378 | -2.40E-20 |
| 595 | 7.23E-03 | 1.45E-19 | -2.95E-20 | -0.01243567048 | -2.38E-19 | 1.33E-19 | 2.85E-19 | -2.24E-20 | 0.02198534022 | 3.46E-19 | -7.70E-20 | 8.89E-19 | 1.76E-19 | -0.0023957340234023 | 0.00E+00 |
| 620 | 8.84E-03 | -4.90E-19 | 4.73E-21 | -0.0158614715 | 5.30E-19 | 4.06E-20 | 5.99E-20 | 2.01E-20 | 0.02657583541 | 2.47E-19 | 5.35E-20 | -1.19E-19 | -2.99E-20 | -0.0032778399978 | -1.50E-20 |
| 645 | 8.87E-03 | -3.64E-19 | -2.47E-20 | -0.01744489829 | 5.92E-19 | 1.12E-19 | 1.46E-19 | 2.80E-20 | 0.02674635366 | 2.74E-19 | 1.05E-20 | 4.07E-21 | 7.82E-21 | -0.0026177999596 | 2.74E-20 |
| 670 | 8.42E-03 | -3.36E-20 | 7.69E-21 | -0.01723340287 | -1.85E-19 | -3.14E-21 | 2.14E-19 | 3.99E-20 | 0.02551973278 | 5.14E-19 | 4.35E-20 | -1.21E-18 | -1.22E-20 | -0.00204955599 | -1.22E-20 |

**Table S14: Mid NA Broadband Lens: [NA = 0.36 (3 um – 5 um)]**

| Wavelength (nm) | Piston | Tip | Tilt | Defocus | Vertical astigmatism | Oblique astigmatism | Horizontal coma | Vertical coma | Primary spherical | Oblique trefoil | Vertical trefoil | Vertical secondary astigmatism | Oblique secondary astigmatism | Vertical quadrafoil | Oblique quadrafoil |
|---|---|---|---|---|---|---|---|---|---|---|---|---|---|---|---|
| 3 | 5.24E-03 | -4.26E-05 | 4.26E-05 | -0.006282194933 | 4.12E-19 | -1.26E-20 | -0.0003431500251 | 3.43E-04 | 0.01353879275 | -9.87E-05 | -9.87E-05 | -1.10E-19 | -3.24E-20 | 0.001569126352 | 3.24E-20 |
| 3.25 | 4.64E-03 | -1.75E-05 | 1.75E-05 | -0.006956088722 | 1.21E-19 | -6.74E-21 | -0.0003399400209 | 3.40E-04 | 0.01301748094 | -5.13E-05 | -5.13E-05 | 2.25E-19 | 8.70E-21 | 0.0006576337018 | 9.57E-20 |
| 3.5 | 4.73E-03 | -1.73E-05 | 1.73E-05 | -0.007551473879 | 6.99E-20 | -1.16E-19 | -0.0003880635223 | 3.88E-04 | 0.01359357705 | -1.16E-05 | -1.16E-05 | 1.42E-19 | 1.03E-19 | 0.0007015665668 | -1.87E-20 |
| 3.75 | 5.22E-03 | 1.32E-05 | -1.32E-05 | -0.008925515569 | 4.38E-19 | -3.89E-21 | -0.0004279703352 | 4.28E-04 | 0.01519448031 | 1.04E- | 1.04E- | 3.56E-20 | -5.03E-21 | 0.0007907630242 | 1.51E-20 |

| | | | | | | | | | | | 05 | 05 | | | | |
|---|---|---|---|---|---|---|---|---|---|---|---|---|---|---|---|---|
| 4 | 6.15E-03 | 4.01E-05 | -4.01E-05 | -0.01081581694 | -5.08E-20 | -5.99E-20 | -2.11E-04 | 2.11E-04 | 0.017385 2022 | -1.33E-05 | -1.33E-05 | 9.19E-20 | 1.60E-20 | 0.0010037 89483 | 1.60E-20 |
| 4.25 | 7.15E-03 | 6.77E-05 | -6.77E-05 | -0.01199737678 | 3.20E-19 | -8.81E-21 | -2.13E-04 | 2.13E-04 | 0.020153 83329 | -3.23E-05 | -3.23E-05 | -7.74E-20 | -5.69E-21 | 0.0011998 22716 | -3.41E-20 |
| 4.5 | 7.14E-03 | 8.81E-05 | -8.81E-05 | -0.01237778106 | -5.92E-20 | -3.75E-20 | -2.00E-04 | 2.00E-04 | 0.021764 61854 | 1.22E-05 | 1.22E-05 | -1.30E-20 | -1.82E-20 | 0.0011773 27913 | 6.05E-21 |
| 4.75 | 6.69E-03 | 3.95E-05 | -3.95E-05 | -0.01266603737 | -2.20E-20 | -2.23E-20 | -0.0003298 267573 | 3.30E-04 | 0.022327 16133 | -4.99E-05 | -4.99E-05 | 1.79E-20 | -3.20E-20 | 0.0009656 442358 | -4.80E-20 |
| 5 | 6.72E-03 | 1.13E-05 | -1.13E-05 | -0.01256855892 | -1.01E-20 | -2.08E-20 | -0.0004088 105373 | 4.09E-04 | 0.022347 37438 | -4.58E-05 | -4.58E-05 | -6.55E-20 | -6.05E-20 | 0.0004061 059614 | 9.41E-20 |

**Table S15: High NA Broadband Lens: [NA = 0.81 (560 nm – 810 nm)]**

| Wavelength (nm) | Piston | Tip | Tilt | Defocus | Vertical astigmatism | Oblique astigmatism | Horizontal coma | Vertical coma | Primary spherical | Oblique trefoil | Vertical trefoil | Vertical secondary astigmatism | Oblique secondary astigmatism | Vertical quadrafoil | Oblique quadrafoil |
|---|---|---|---|---|---|---|---|---|---|---|---|---|---|---|---|
| 560 | 1.85E-02 | 1.84E-19 | -6.75E-20 | -0.024760 71025 | 3.10E-19 | 8.26E-20 | 2.04E-19 | -9.55E-20 | 0.040656 20221 | 1.65E-18 | -4.77E-20 | 5.20E-19 | 2.13E-19 | 0.0040490 14203 | 0.00E+00 |
| 585 | 1.86E-02 | -2.60E-20 | 0.00E+00 | -0.025344 8257 | 1.11E-18 | 8.65E-20 | 6.55E-19 | 1.00E-19 | 0.038980 29116 | 1.69E-18 | 0 | -7.25E-19 | -1.12E-19 | 0.0050142 91195 | 2.23E-19 |
| 610 | 1.87E-02 | 2.00E-19 | -7.39E-20 | -0.028762 55086 | 5.03E-19 | 0.00E+00 | 3.82E-19 | 0.00E+00 | 0.038212 77303 | 1.84E-18 | -5.23E-20 | -4.30E-19 | 0.00E+00 | 0.0026347 13919 | -1.75E-19 |
| 635 | 2.01E-02 | -5.12E-19 | 3.84E-20 | -0.029578 44367 | -1.22E-18 | 9.39E-20 | 2.96E-19 | 1.63E-19 | 0.036954 33521 | 1.48E-18 | -5.43E-20 | 4.45E-19 | 1.21E-19 | 0.0004871 430698 | 6.06E-20 |
| 660 | 2.11E-02 | -4.36E-19 | 0.00E+00 | -0.030394 7465 | -2.50E-19 | 9.78E-20 | 9.45E-19 | 0.00E+00 | 0.034618 92415 | 1.03E-18 | 1.13E-19 | 1.94E-19 | 0.00E+00 | 7.96E-05 | 0.00E+00 |
| 685 | 2.16E-02 | -5.61E-19 | -2.08E-20 | -0.032665 87241 | 2.97E-18 | 7.66E-20 | 7.65E-19 | -1.18E-19 | 0.031866 84042 | 1.38E-18 | 5.90E-20 | -5.48E-19 | 0.00E+00 | 1.64E-04 | 1.65E-19 |
| 710 | 2.46E-02 | 7.70E-19 | -1.29E-19 | -0.033940 94395 | 1.25E-19 | 0.00E+00 | 1.16E-18 | 0.00E+00 | 0.032888 78716 | 1.77E-18 | -1.22E-19 | -2.31E-18 | -1.36E-19 | 0.0026186 02593 | 6.80E-20 |
| 735 | 2.77E-02 | -1.50E-18 | 4.50E-20 | -0.034818 05191 | 1.53E-18 | 5.51E-20 | 7.64E-19 | 1.27E-19 | 0.030481 89555 | 2.04E-18 | -1.91E-19 | 1.30E-19 | -1.42E-19 | 0.0052473 69014 | 1.42E-19 |
| 760 | 3.05E- | -2.9 | 1.17E- | -0.038873 | -7.91E- | 2.29E-19 | 1.70E-18 | 3.31E-20 | 0.030349 68533 | 2.53E- | -6.61E | -8.62E- | -2.96E- | 0.0031198 67851 | 7.38E-20 |

|  | 02 | 6E-21 | -19 | 01251 | 19 |  |  |  |  | 18 | -20 | 19 | 19 |  |  |
| --- | --- | --- | --- | --- | --- | --- | --- | --- | --- | --- | --- | --- | --- | --- | --- |
| 785 | 3.31E-02 | -1.25E-18 | 0.00E+00 | -0.04036800884 | 3.11E-18 | 0 | 2.84E-18 | 1.35E-19 | 0.02785844987 | 1.61E-18 | -6.77E-20 | -8.63E-19 | -2.27E-19 | 0.001493326151 | 1.51E-19 |
| 810 | 4.23E-02 | -4.21E-19 | 4.96E-20 | -0.04563254646 | -2.68E-18 | 6.08E-20 | 1.90E-18 | 2.11E-19 | 0.02212167989 | 3.57E-18 | 0.00E+00 | -1.88E-18 | 7.85E-20 | 0.0007521237151 | 0.00E+00 |